\documentclass[twocolumn,twoside,slac_two]{revtex4}
\usepackage{graphicx}
\usepackage{fancyhdr}
\renewcommand{\headrulewidth}{0pt}
\renewcommand{\footrulewidth}{0pt}

\def\be{\begin{equation}}
\def\ee{\end{equation}}

\begin{document}

\title{Emission From Rotation-Powered Pulsars: Is It All Relative?}

%

\author{Alice K. Harding}
\affiliation{NASA Goddard Space Flight Center, Greenbelt, MD 20771, USA}

\begin{abstract}
Thirty-five years after the discovery of rotation-powered pulsars, we still 
do not understand the fundamentals of their pulsed emission at any wavelength.  
Even detailed pulse profiles cannot identify the location of the 
emission in a magnetosphere that extends from the neutron star surface to near 
the light cylinder.  Compounding the problem are effects of strong gravity 
at low altitudes and plasma moving at relativistic speeds in the outer magnetosphere.  
I will discuss the role of special and general relativistic effects on pulsar 
emission, from inertial frame-dragging near the stellar surface to aberration, 
time-of-flight and retardation of the magnetic field near the light cylinder.    
Understanding how these effects determine what we observe at different wavelengths 
is critical to unraveling the emission physics
\end{abstract}

\maketitle

\thispagestyle{fancy}


\section{INTRODUCTION}

Rotation-powered pulsars shine over a broad spectrum from radio to high-energy 
$\gamma$-ray wavelengths.  Pulsed emission in the radio band, where they were first 
discovered, has been detected from over 1500 objects and the characteristics of this
emission have been studied in great detail for many years.  Even so, the origin of the radio
pulsation is still not understood, except to realize that it must be a coherent process
requiring significant particle densities and probably electron-positron pairs \cite{Melrose00}.
Studies of radio pulse morphology and phase-resolved polarization patterns suggest that
the emission is radiated along open dipole magnetic field lines within several hundred
stellar radii of the polar cap \cite{Rankin93, KG03}.  
Pulsed emission in optical, X-ray and $\gamma$-ray bands 
has been detected from a much smaller
number of sources.  Some thirty pulsars are known to have pulsations at X-ray energies \cite{KRH04}, 
ten in the optical band \cite{Mignani04} and about ten at $\gamma$-ray energies \cite{Kanbach02}.
Since emission at the higher energies originates from incoherent processes, there is greater
hope of understanding its origin, as well as the particle acceleration processes in the 
magnetosphere.  But nature seems to be making things difficult for us here as well, as in most 
cases the high-energy pulse profiles do not look like the radio profiles and the phases
of the pulses are different across energy bands.  There are, however, notable exceptions to
this behavior.  The Crab pulsar exhibits a double-peaked pulse profile that is very similar
at all wavelengths from radio to high-energy $\gamma$ rays, and the pulses are in phase
across the entire spectrum to within tenths of milliseconds \cite{Kuiper03}.  Several other pulsars
with comparably short or shorter pulse periods show similar behavior, having X-ray and
radio profiles that resemble each other and are in phase.  These sources include PSR B0540-69,
a 50 ms Crab-like pulsar in the LMC, PSR J1617-5055, a 69 ms pulsar in RCW103, and the millisecond
pulsars PSR B1821-24, PSR J0437-4715 and PSR B1937+21.  Perhaps this indicates that the emission
at these different wavelengths originates in the same small region of the magnetosphere.  
However, there is another possibility if one takes into account relativistic effects on 
emission, and which may agree better with emission models as well as constraints on viewing
angles and polarization measurements.

The key to deciphering pulsar emission, I believe, is to understand the relativistic effects
that operate near a rapidly spinning, magnetized star with strong gravity, to build these into
our models, and to recognize how they distort the radiation we observe.  If we can learn to
deconvolve these effects from the signals, or to identify signatures which can give us information 
on the location of the emission, then perhaps we will have clues to the acceleration and emission 
mechanisms.  Special relativistic effects such as aberration, time-of-flight delays and 
retardation of the magnetic field become important for acceleration and radiation in the
outer magnetosphere.  General relativistic effects such as inertial frame-dragging,
gravitational red-shift and light bending are important nearer the neutron star surface.
I will review these special and general relativistic effects in pulsar magnetospheres,
how they are being incorporated into pulse emission models and how they are essential to
accurately interpreting pulsar data to determine the geometry of the emission.  

\section{SPECIAL RELATIVISTIC EFFECTS} \label{sec:SR}

Pulsars rotate rapidly enough that the star's exterior magnetic field and charges that are
coupled to it will experience relativistic corotation velocities in the outer parts of the
magnetosphere.  Since the corotation velocity at radius $r$, for angle $\zeta$ to the rotation axis,
is $\beta_{rot} = \Omega r\sin\zeta/c$, first order special relativistic effects become important when
$r$ is a significant fraction of the light cylinder radius, $R_{LC} = c/\Omega$, where
$\Omega$ is the pulsar rotation frequency.   Aberration causes the photon emission directions
to appear shifted to a non-rotating observer in the direction of rotation, so the emission
arrives at an earlier phase.  The first order phase shift due to aberration, at an emission 
radius $r_{\rm em}$, is \cite{GG01, DRH04}
\be  \label{phi_ab}
\Delta\phi_{\rm ab} \simeq -{r_{\rm em} \over R_{LC}},
\ee
which is independent of the angle $\zeta$.
In addition, there are time-of-flight delays in the arrival time of radiation emitted at 
different $r_{\rm em}$.  The phase delay is, to first order in $\beta_{rot}$, the same as that
due to aberration, or
\be   \label{phi_ret}
\Delta \phi _{\rm ret}  \approx  - \frac{{r_{\rm em} }}{{R_{LC} }}.
\ee
 
\begin{figure}[t]  
\hskip -83pt
\includegraphics[width=11cm]{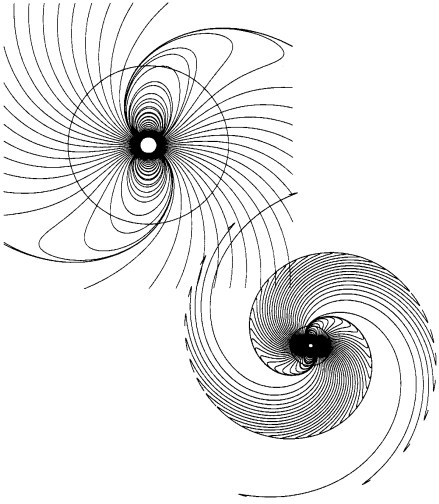}
\caption{Sweepback of the vacuum dipole magnetic field in the spin-equatorial plane (from 
 \cite{Yad97}) for pulsar inclination angle is $60^{\circ}$.  Rotation is counterclockwise and 
the circle denotes the light cylinder radius.  Top view shows inside and just outside the light cylinder
and bottom view shows the field forming a spiral pattern far outside the light cylinder.} \label{f1}
\end{figure}

The magnetic field of the neutron star will be distorted by retardation in the outer parts
of the magnetosphere near the light cylinder.  The original Deutsch \cite{Deutsch55} solution for a
rotating dipole field in vacuum exhibits such an effect, as the `near field' dipole must match
to the `far-field' electromagnetic wave solution.  Beyond the light cylinder the field 
lines become swept back, 
in the opposite direction to that of rotation (see Figure 1).  However, the field begins to be 
distorted within the magnetosphere, and this can affect both the direction of emission tangent 
to field lines as well as the shape of the open field volume \cite{Yad97, AE98}.  
The lowest order change in the
field direction caused by the sweep-back inside the light cylinder is \cite{Shitov83} 
\be
\delta_{\rm sb} \simeq 1.2\left(\frac{{r_{\rm em} }}{{R_{LC} }}\right)^3 \sin^2\alpha,
\ee
where $\alpha$ is the magnetic inclination angle, which is insignificant compared with 
the phase shifts due to aberration and time-of-flight.
There is however a much more important distortion of the open field volume, due to field-line 
sweep-back near the light cylinder, which results in distortion and displacement of the polar cap 
at the surface of the neutron star.  The displacement of the open field volume causes a phase
shift \cite{DH04} 
\be  \label{phi_ov}
\Delta \phi _{\rm ov} \simeq 0.2\left(\frac{{r_{\rm em} }}{{R_{LC} }}\right)^{1/2}
\ee
which is of lower order than aberration and time-of-flight delay and thus will dominate
at smaller emission radii.  The retarded vacuum dipole solution has been widely used in modeling
pulsar emission \cite{RY95, CRZ00, Shitov83}, even though active
pulsar magnetospheres will contain charges and currents, because solutions for the realistic case
are more complex and model dependent.  An analytic solution for a pulsar magnetosphere with 
currents on open field lines in a space-charge limited flow (SCLF) model \cite{MH05} illustrates 
that there are important 
differences in the vacuum and non-vacuum cases.  For example, there is no sweepback of field lines 
for an aligned rotator in the vacuum case (where there is no spin-down torque) but there {\it is}
sweepback in the case of the non-vacuum aligned rotator due to the current flow.

\subsection{GENERAL RELATIVISTIC EFFECTS} \label{sec:GR}

\begin{figure*}[t]
\includegraphics[width=120mm]{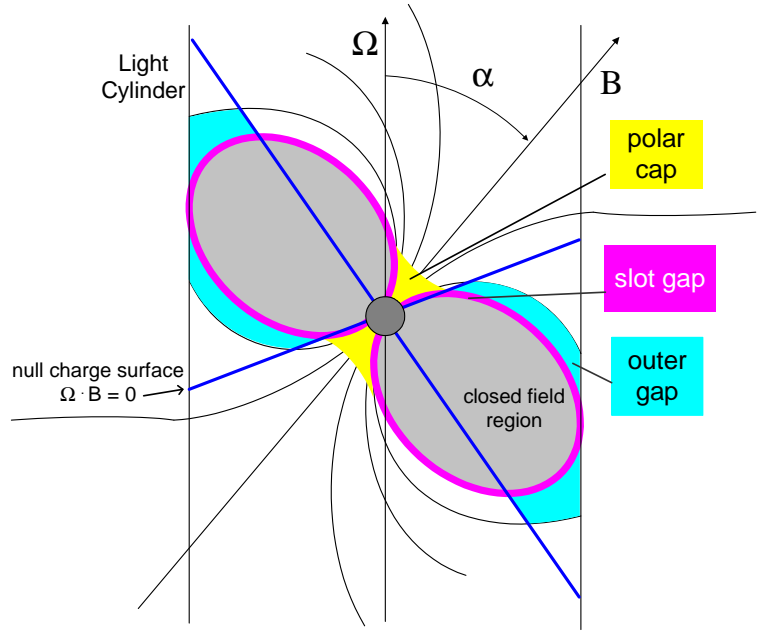}
\vskip -6.0cm
\caption{Schematic illustration of pulsar acceleration/emission models.} \label{f2}
\end{figure*}

Distortions of space-time near the neutron star produce various effects important to pulsar 
radiation, including gravitational red-shifting of photons, bending of light, curved space-time 
changes to the magnetic field and inertial frame-dragging.  The frequency of radiation emitted
near the neutron star surface is red-shifted as photons climb out of the gravitational potential
well toward a distant observer in the (relatively) flat space-time of Earth's gravity.  The
resulting decrease in energy of the photons emitted at radius $r_{\rm em}$  is
\be
\varepsilon  = \varepsilon '\left( {1 - \frac{{2GM}}{{r_{\rm em}c^2 }}} \right)^{1/2}. 
\ee
Such a red-shift can also have secondary effects on the radiation, such as an increased efficiency
of attenuation by one-photon pair creation \cite{GH94} and photon splitting \cite{BH01}, 
since these processes depend sensitively on photon energy.  

Gravitational bending of photon paths is most pronounced for trajectories at large angles to the 
radial direction.  For non-thermal radiation by highly relativistic 
particles that is emitted mostly tangent to magnetic dipole field lines in polar regions, light
bending has a second-order effect on radiation direction and attenuation.  However, for thermal
emission from a hot neutron star surface, emitted semi-isotropically, light bending has a dominant 
effect on smearing-out pulse profiles and
diminishing degree of modulation \cite{Page95, RM88}.

The dipole field in a Schwarzschild metric \cite{WS83} exhibits several features
that are important for pulsar radiation.  The surface field strength at the
magnetic poles is increased, relative to flat space-time, so that
\be
B_0^{GR}  \approx (1 + \frac{{r_g }}{R})_{} B_0^{Flat} 
\ee
where $R$ is the radius and $r_g$ is the gravitational radius of the neutron star.  
This effect also enhances the
photon attenuation due to pair creation and photon splitting.  The dipole field is also less
open, as field lines near the light cylinder have footpoints that are nearer the magnetic axis.
There is a resulting decrease in polar cap half-angle,
\be
\theta _{GR}^{}  \approx \frac{{\theta _{Flat} }}{{\sqrt {1 + r_g /R} }}.
\ee
Effectively though, for emission of photons tangent to field lines, the effect of light bending 
tends to cancel the effect of the smaller polar cap size, so that the resulting size of the
emission beam is roughly the same as in flat space-time \cite{GH94}.

The fast rotation of the neutron star also distorts the nearby space-time, resulting in a 
dragging of the inertial frame around it (also known as the Lense-Thirring effect).  
In the Kerr metric that includes the rotation of the neutron star, the inertial frame rotates
with angular velocity, $\omega  = {2G{\rm{L}}}/{c^2 r^3 }$, if $L$ is the neutron star angular momentum, 
which is effectively a
differential rotation.  If one solves Maxwell's equations in this metric to obtain the electric
field induced by the rotation of the magnetic field \cite{MT92}, there 
is a correction to the Goldreich-Julian charge density (the charge density required to screen 
the electric field parallel to the magnetic field),
\be
\rho _{GJ}  \approx  - \frac{1}{{4\pi c}}\nabla \cdot \left[ {\frac{{\rm{1}}}{\alpha }
{\rm\bf {(v - \omega)}} \times B} \right],
\ee
where ${\bf v = \Omega \times r}$.  The inertial-frame angular velocity now appears in the expression
for $\rho _{GJ}$, as well as the rotational velocity ${\rm\bf v}$.  
This change in $\rho _{GJ}$ turns out to have a significant effect for SCLF 
models \cite{AS79}, where the accelerating electric field results
from the small difference between the actual charge flow along the open field lines and the GJ charge flow,
\[
\nabla \cdot E_{||}  = 4\pi (\rho  - \rho _{GJ} )
\]
The resulting parallel electric field near the neutron star is 10 to 50 times higher that the
corresponding field in flat space-time (MT92), so that the frame-dragging contribution actually
dominates.  The frame-dragging $E_{\parallel}$ also enables particle acceleration for aligned
rotators and along field lines curving away from the rotation axis, which does not occur in flat
space-time.    

\section{DECODING THE SIGNALS}

\begin{figure}[t]
\hskip -2cm
\includegraphics[width=110mm]{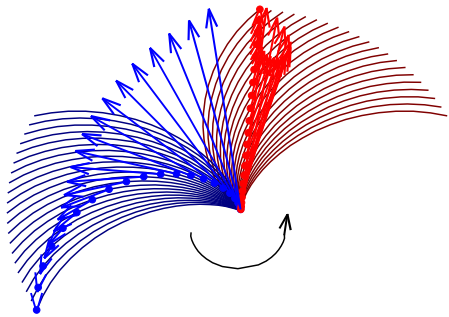}
\vskip -3cm
\caption{Time-lapse illustration of formation of caustic emission in the spin-equatorial plane of an
orthogonal rotator.  The magnetic axis is vertical at time $t = 0$ and rotates counterclockwise.  
Dots and arrows mark the successive emission points and directions of photons along the leading and
trailing last open field lines that will arrive 
at a distant observer simultaneously.  
 } \label{f3}
\end{figure}

Many of the effects described above have been incorporated in pulsar emission models and in some
cases, they are essential features.  Most current models for high energy emission involve some particle
acceleration and radiation in the outer magnetosphere.  Although traditional polar cap 
models \cite{DH96, HM98} focus on activity near the neutron star, 
more recently such models have explored extended acceleration from the neutron star surface to high 
altitudes in the slot gap along the edge of the open field region
\cite{MH04} (see Figure 2).  
Acceleration in outer gap models \cite{RY95, CRZ00, Hir01} occurs in the outer magnetosphere
between the null charge surface and the light cylinder along the last open field line, 
although recent 2D solutions of the gap geometry show that the gap may extend to the inner magnetosphere 
\cite{TSH04}.  
Polar cap and outer gap model geometry thus seems to be merging, although the electrodynamics of the
two models remains fundamentally different.  In both pictures, special relativistic effects are essential
in modeling the high altitude emission.  General relativistic effects are additionally important in 
polar cap models.

\subsection{Caustics} \label{sec:caustic}

A curious feature of the emission pattern in the outer magnetosphere of a rotating dipole
was first noted by Morini \cite{Morini83}.  If one assumes that photons are radiated tangent to the magnetic
field from the polar cap to the light cylinder, 
then the relative phase shifts of photons emitted at 
different radii due to dipole curvature, aberration and time-of-flight nearly cancel on  
field lines on the trailing edge of the open region (see Figure 3).  Radiation along such trailing field lines 
bunches in phase, forming a sharp peak in 
the profile.  On the other hand, photons emitted at different radii along leading field lines spreads 
out in phase.  The effect is most pronounced for large inclination of the magnetic axis to the rotation axis.
A plot of observer angle to the rotation axis versus phase, as shown in Figure 4a, clearly displays the
sharp lines of emission, or caustics, in the radiation from trailing field lines.  A number of emission
models \cite{CHR86, Smith86, RY95, DR03} have made use of the caustic effect to produce the 
sharp peaks seen in profiles of high-energy pulsars.  In purely geometrical schemes, the high energy emission 
could originate from caustics associated with either one magnetic pole \cite{Morini83, Smith86, RY95}, 
or both poles \cite{DR03}.   Two types of physical acceleration and emission models divide in this way as well, 
with outer gap models exhibiting one-pole caustic geometry and slot gap models exhibiting two-pole
caustics.

\begin{figure*}[t]
\vskip -0.5cm
\includegraphics[width=125mm]{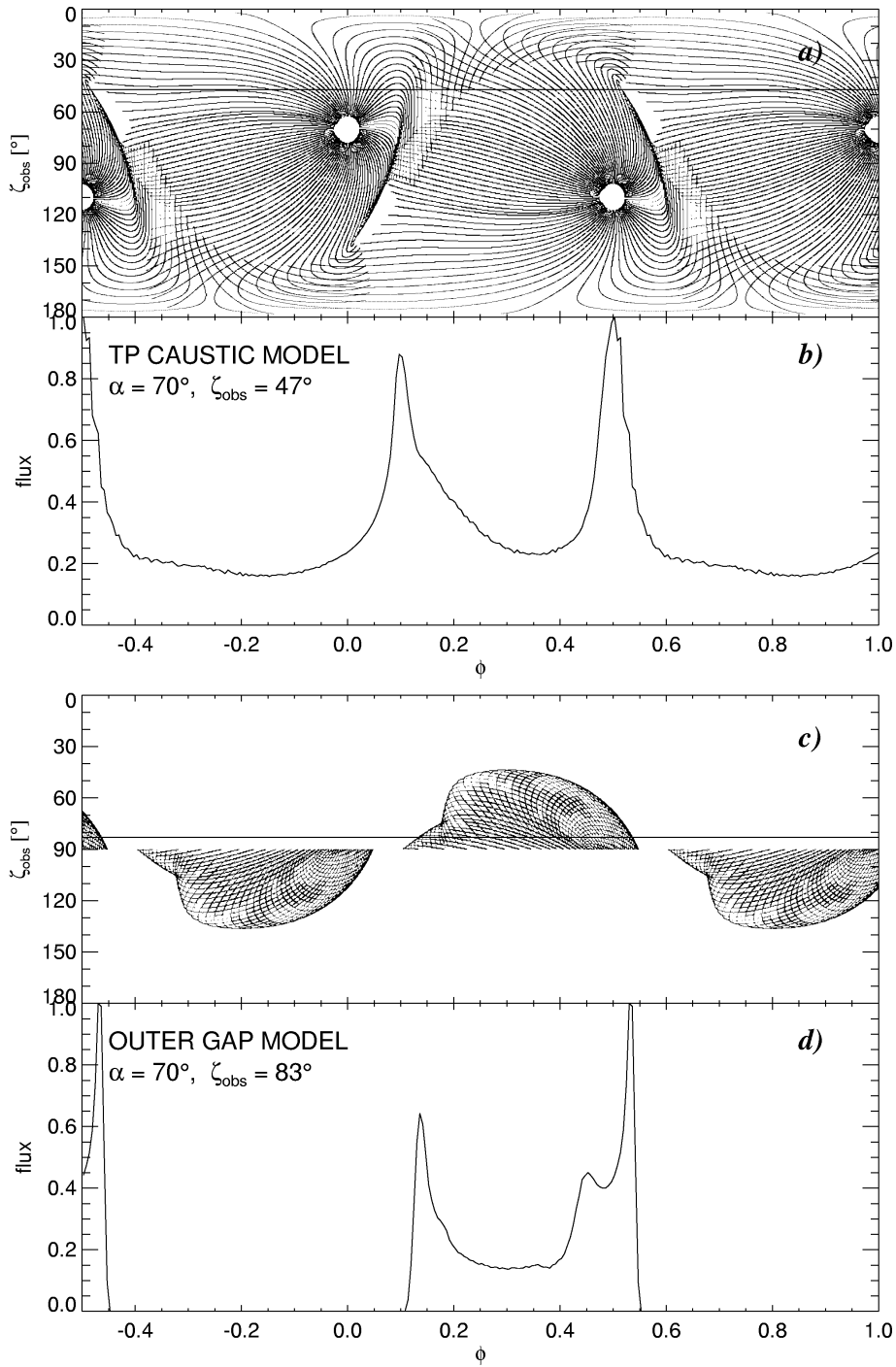}
\caption{Comparison of two-pole caustic and outer gap models. 
a) Plot of emission in the ($\zeta_{\rm obs}$, $\phi$) plane, calculated with a two-pole caustic
model for inclination, $\alpha = 70^{\circ}$.  $\phi$ is the rotational phase and $\zeta_{\rm obs}$ is the 
viewing angle measured from the rotation axis. b) High-energy profile for $\zeta_{\rm obs} = 47^{\circ}$, 
produced by a horizontal cut through the phase plot above.  c) As in a) for an outer gap model.
d) High-energy profile from the outer gap phase plot c) for $\zeta_{\rm obs} = 83^{\circ}$.
} \label{f4}
\end{figure*}

\subsection{Outer Gap Model}

Outer-gap models  \cite{CHR86,Rom86} assume that acceleration occurs in vacuum gaps that 
develop in the outer magnetosphere, along the last open field line above the null charge surfaces, 
where the Goldreich-Julian charge density changes sign (see Figure 2), and that 
high-energy emission results from photon-photon pair production-induced cascades.  The pair
cascades screen the accelerating electric field and limit the size of the gap both along and 
across the magnetic field.  The geometry
of the outer gaps prevents an observer from seeing gaps associated with both magnetic poles, putting
them in the class of one-pole caustic models.  The high energy pulse profiles that are observed to 
have two widely separated, sharp peaks are formed by caustics, the leading peak originating from 
overlapping field lines at $r_{\rm em} \sim 0.9\,R_{LC}$ and the trailing peak originating from the 
caustic along trailing field lines at $r_{\rm em} \sim 0.2 - 0.8\,R_{LC}$.  An example of outer gap 
profile formation is shown in Figure 4c and 4d.  A drawback to this
type of model is that the formation of the leading peak is very sensitive to the structure of the
retarded field lines very near the light cylinder, which is not well known for non-vacuum models (see
Section \ref{sec:SR}). Some recent outer gap models point out that the gap lower boundary should exist
somewhat below the null charge surface, and the location of this boundary will depend on the external 
currents that flow into or out of the gap \cite{HSH03, TSH04}.  If the gap moves close enough to the
neutron star, depending on inclination angle, it might be possible for an observer to see gaps from both
magnetic poles.

\subsection{Slot Gap Model}

The slot gap, a narrow pair-free accelerator bordering the closed field region, is a feature 
of polar cap SCLF models (AS79).  SCLF models assume freely emitted charges flow out from the 
neutron polar cap along open field lines.  Since this charge flow is not sufficient to supply the
Goldreich-Julian charge above the surface, as discussed in Section \ref{sec:GR}, 
an $E_{\parallel}$ exists and charges are accelerated.
Radiation from these charges forms electron-positron pairs in the strong magnetic field, which can screen
the $E_{\parallel}$ above a pair front in a distance small compared to the acceleration distance.
These models assume a boundary condition that the accelerating electric field and potential vanish 
at the last open field line.  Near the boundary, the electric field is decreasing and a larger distance is 
required for the electrons to accelerate to the Lorentz factor needed to radiate photons energetic 
enough to produce pairs.  The pair front thus occurs at higher and higher altitudes as the boundary 
is approached and curves upward, becoming asymptotically parallel to the 
last open field line.  Since $E_{\parallel}$ is unscreened in the slot gap, particles continue to
accelerate and radiate to high altitude along the last open field lines.  It is interesting that 
frame-dragging's dominant effect on the accelerating field persists even at large distances from the
neutron star surface, since the high-altitude SCLF solution depends on surface 
boundary conditions \cite{MH04}.  The resulting emission
geometry naturally creates caustics visible from both poles, and so produces double-peaked
pulse profiles similar to the two-pole caustic model of Dyks \& Rudak \cite{DR03}, as shown in Figures 
4a and 4b.  
In contrast to one-pole outer gap models, slot gap models have outward emission beams below
the null-charge surface.  Both peaks originate from trailing field-line caustics at intermediate 
radii $r_{\rm em} \sim 0.1 - 0.7\,R_{LC}$.   In such profiles, the leading `outer gap' peak appears
as a bump just following the first peak.   

\subsection{Comparing Outer Gap and Slot Gap Models}

Although outer gap and slot gap models can produce similar pulse profiles, phase plots of their emission 
show prominent differences.  As displayed in Figure 4a, slot gap emission fills the entire sky even
though radiation is produced only on field lines bordering the closed regions.  Since the radiation is 
emitted over the entire length of these field lines from the magnetic poles (the hollow circles in Fig. 4a)
to near the light cylinder (where field lines from opposite hemispheres overlap), observers can view the
emission from both magnetic poles at a wide range of angles.  Two-pole caustic emission can be seen for
inclination angles $\alpha \gtrsim 30^{\circ}$ \cite{DHR04}.  Because the emission fills the sky, radiation
will be visible at all pulse phases so that pulse profiles like the one shown in Figure 4b will
include `off-pulse' emission.  The bump or shoulder appearing on the inside of the first peak is due
to emission from overlapping field lines from opposite poles, and occurs at the same phase as the
first peak in outer gap models.  It is interesting that the $\gamma$-ray profile of the 
Vela pulsar displays a similar broad bump \cite{Kanbach02}.

Outer gap emission fills only a fraction of the sky, as shown in Figure 4c, since outward-going radiation
occurs only above the null charge surface in each hemisphere.  The outer gap emission pattern is thus a
subset of the two-pole caustic emission pattern.  In Figure 4c, the magnetic poles are not visible and
part of the caustic near each pole is cut off.  The pattern fills the most phase space for large inclination
angles, but shrinks to a small area near the rotational equator for small inclination \cite{CRZ00}.  
Outer gap emission is never visible at small viewing angles $\zeta_{\rm obs} \lesssim 30^{\circ}$ at 
any inclination angle.
An observer cutting through the pattern at a single $\zeta_{\rm obs}$
will see emission only from field lines originating at one pole.  As a result, the peaks in outer gap 
profiles drop sharply at their outer edges and there is no off-pulse emission outside the peaks, 
as is evident in
Figure 4d.  The second peak has the same origin as the second peak in the two-pole caustic model, 
but the first peak is formed by emission from overlapping field lines very near the light cylinder 
(similar to the formation of the bump in Figure 4b).

\section{DETECTING THE SIGNATURES}

\begin{figure*}[t]
\includegraphics[width=120mm]{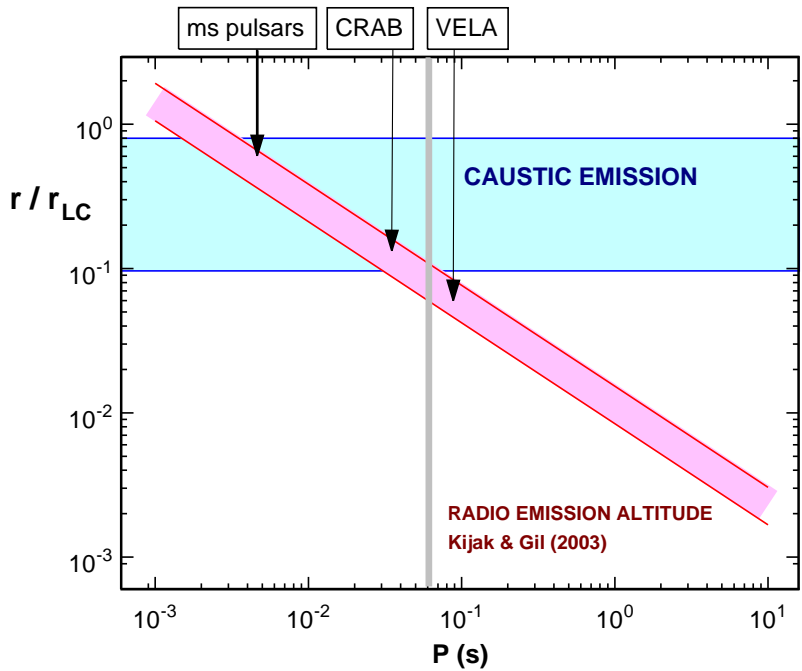}
\caption{Radio emission radius (in units of light cylinder radius) as a function of pulsar period
from  \cite{KG03}.  Also shown is the range of radii there caustic emission occurs.
} \label{f5}
\end{figure*}

Emission models can make predictions of certain observable signatures, which may provide clues to both
the relativistic effects that are shaping radiation patterns and to the location and nature of
the radiation.   Such signatures are found in the relative phases of emission at different wavelengths,
the polarization patterns and the spectrum of radiation.

\subsection{Phase Shifts} \label{sec:phase}

Multiwavelength behavior of observed pulse profiles contains a wealth of information, but has proven 
to be extremely complex and difficult to interpret.   With detailed modeling of emission, taking into
account effects of rotation and strong gravity of the neutron star, we can possibly understand why many
of the short period pulsars have pulse profiles that show phase coherence of emission at different
wavelengths while longer period pulsars do not.  For example, as discussed in Section \ref{sec:caustic},
caustic emission along trailing field lines puts emission radiated over a large range of different altitudes
in phase for a distant observer.  This means that any radiation originating between 
$r_{\rm em} \sim 0.1\,R_{LC}$ and $r_{\rm em} \sim 0.8\,R_{LC}$ near the last open field line 
will appear in phase, but radiation
originating outside this region will appear at different phases.  This feature of caustics may offer 
an explanation of multiwavelength phase coherence seen in the profiles of fast pulsars.  The altitude of 
caustic formation is the same fraction of light cylinder radius in all pulsars, independent of period. 
However, emission at various wavelengths in different pulsars may not all originate within the radii of
caustic formation.  To illustrate this point, Figure 5 displays the dependence of pulsar radio emission 
altitude as a function of
pulse period, determined empirically from observed pulse widths and radius-to-frequency mapping \cite{KG03}, superposed on the altitude range of caustic formation.  Although the exact altitudes
of pulsar radio emission are not necessarily known this well, there is indication of 
such a dependence on period.
In that case, it can be seen that radio emission will originate within the caustic formation region only 
for the fastest pulsars.  The pattern of radio conal emission, that may original near the last open 
field lines, will then become severely distorted into caustic emission peaks.  
In this example, the dividing point occurs around a period of 60 ms, so that 
radio and high-energy emission would be phase coherent for the Crab and most millisecond pulsars, but not
for Vela and pulsars with longer periods.  This in fact seems to match observation.  The radio emission
in slower pulsars in this picture occurs at altitudes below the caustic region, so that the observer
light-of-sight would cross the radio core and/or conal beams close to the magnetic pole before crossing
the high energy caustics.  

Another example of relativistic effects that cause phase-shift signatures in pulsar emission occurs 
in observed radio profiles.  In pulsars with both core and one or more conal beams (i.e. profiles 
containing three or five peaks), there is often a phase shift between the central core component and the
center of the profile as determined by the conal beam(s).  In most cases, the conal components lead
the core component in phase.  Such phase shifts have been interpreted as due to a combination of 
aberration (Eqn \ref{phi_ab}) and time delay (Eqn \ref{phi_ret}) if the conal emission occurs 
at a higher altitude than the core emission \cite{GG01}.  
Shifts in the center of the polarization position angle curve are also
observed and could be caused by the same effects \cite{BCW91}.  There are some pulsars
whose radio profiles show phase shifts in the opposite direction (the conal components lag the core
components).  This may be due to a dominating phase shift caused by the distortion of the open field
volume due to sweepback of the magnetic field \cite{DH04} (Eqn.  \ref{phi_ov}).
Such phase shift in radio profiles have been used to derive the radio emission altitudes and they
generally agree with those derived by other methods \cite{DRH04}. 

\begin{figure*}[t]
\hskip -0.5cm
\vskip -10.0cm
\includegraphics[width=180mm]{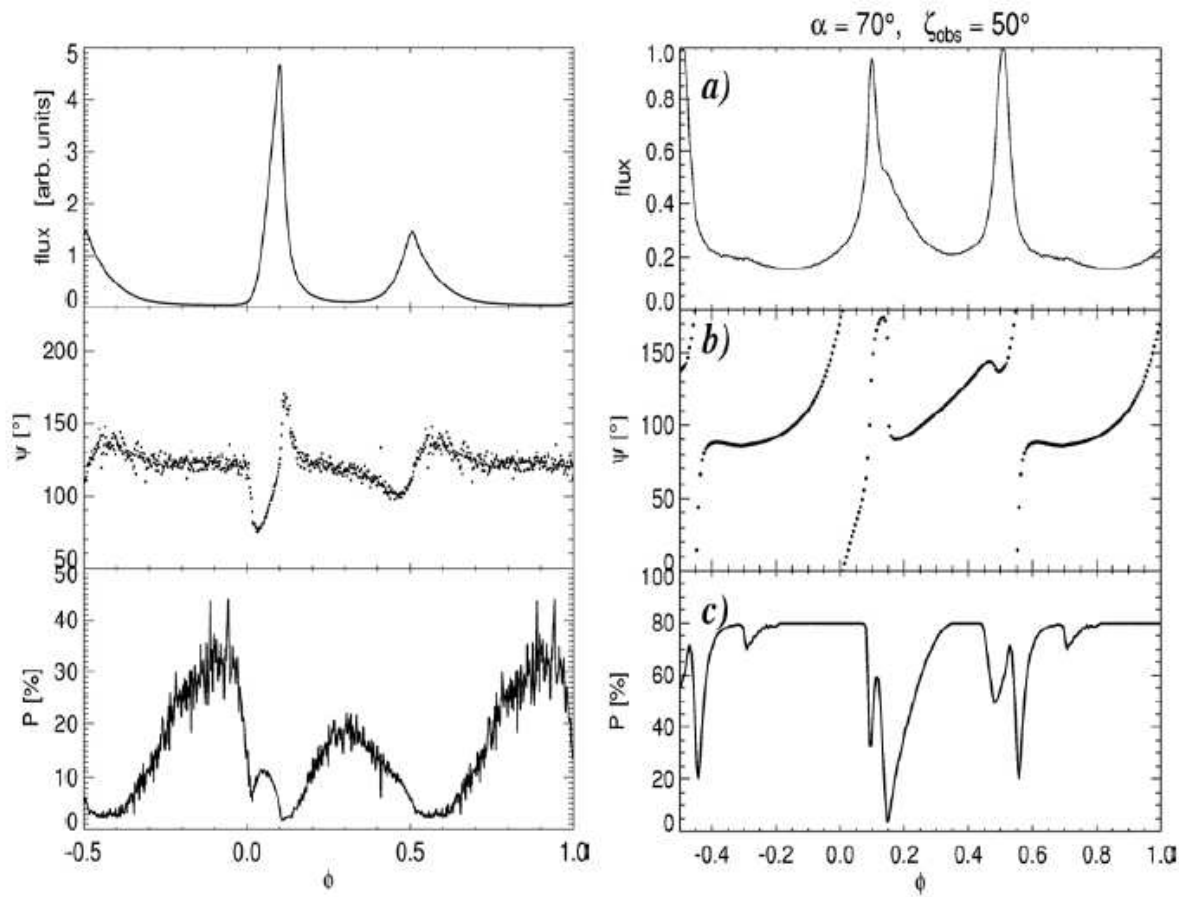}
\caption{Left: Optical profile (top), polarization position angle $\Psi$ (middle) and degree of polarization 
(bottom) of the Crab pulsar from OPTIMA  \cite{Kellner02}.  Right: Model profile and polarization
characteristics predicted by a two-pole caustic model  \cite{DHR04} with dipole inclination 
$\alpha = 70^{\circ}$ and viewing angle $\zeta_{\rm obs} = 50^{\circ}$.  The PA curve has $180^{\circ}$
wraparound ambiguity.
} \label{f6}
\end{figure*}

\subsection{Polarization}

Phase-resolved polarimetry of pulsar emission has proven to be a powerful diagnostic 
at radio wavelengths.
The pulsed non-thermal radiation from relativistic particles in the magnetosphere is tightly 
beamed along the neutron-star magnetic field lines and thus the emitted radiation is believed 
to be highly polarized either parallel or perpendicular to the field lines.  Measurement of 
the polarization properties as a function of pulse phase can provide a 
multidimensional mapping of the field pattern at the emission sites.  
In the classical rotating vector model (RVM) 
 \citep{RC69}, the expected signature of emission near the poles 
of a dipole field, an `S'-shaped swing of the polarization position angle through the pulse 
profile, has been seen from many radio pulsars and has generally been taken as proof that 
the radio emission originates from the open field lines of a magnetic dipole.  However, 
while this classical picture can provide a good measurement of the observer angle with respect
to the magnetic pole, $\beta$, for low-altitude emission, it does not provide an unambiguous
measure of pulsar inclination angle $\alpha$.  Furthermore, if the emission occurs at altitudes
that are more than a few tenths of the light cylinder radius, relativistic effects can
significantly distort the simple polarization characteristics predicted by the classical model.
But such distortions can be turned to our advantage if we can model them well enough to 
understand the signatures of the effects.  

As discussed in Section \ref{sec:phase}, one would expect severe distortion of emission
patterns by special relativistic effects for radiation originating within the caustic emission region
shown in Figure 5.  The emission from fast pulsars would thus be more likely to show signatures
of such effects in their polarization patterns, especially at high energies. 
It would be very important to have polarization data at high energies to explore characteristics
of emission in the outer magnetosphere.
Unfortunately, polarization measurements at wavelengths other than radio exist for very few pulsars. 
In the case of the Crab pulsar, polarization measurements have been made at radio and optical 
wavelengths and do not show the simple shape predicted by the classical RVM.  As shown in 
Figure 6, each peak shows a very sharp swing of position angle (PA) and a drop in degree of 
polarization.  Such PA swings could be interpreted as part of an S-curve for small impact
angle $\beta$, but then one would expect a peak separation of $180^\circ$ (rather than the $140^{\circ}$ 
observed) in a 2-pole, near-surface
emission model, and no drop in degree of polarization.  Figure 6 also shows a computed pulse profile,
PA curve and degree of polarization as a function of phase predicted in a 2-pole caustic model \cite{DHR04},  described in Section \ref{sec:caustic}.  Sharp swings of PA and drops in degree
of polarization are signatures of caustic emission and the qualitative similarity
of the observed and model polarization characteristics seem to strengthen the case for caustic
peaks in the Crab pulsar.  This picture is also consistent with the mutliwavelength phase coherence
property expected for fast pulsars, discussed in Section \ref{sec:phase}.

Optical polarization measurements have recently been made \cite{Kern03} for the slower 384 ms 
pulsar PSR B0656+14.  
The double-peaked optical profile is not in phase with the single radio
peak, but the classical RVM provides a consistent fit of both radio and
optical PA data.  One interpretation of these results is that both radio and optical emission
from this middle-aged pulsar originate at low altitudes, with the high-energy emission described by
a hollow-cone centered on the magnetic pole and the radio core emission, as predicted by 
the traditional polar cap model \cite{DH96}.  

\subsection{Spectra}

Spectral signatures can also provide clues to emission altitudes as well as some relativistic effects.
Polar cap \cite{DH96} and slot gap \cite{MH03} models include a component of emission from low-altitude 
pair cascades.  The observable spectra of the cascade emission will exhibit sharp high-energy
cutoffs due to pair attenuation, at an energy of approximately \cite{BH01}
\be \label{Ec}
E_c \sim 2\,\,{\rm GeV}\,P^{1/2}\,\left({r\over R}\right)^{1/2}\, {\rm max}
\left\{0.1, \,B_{0,12}^{-1}\,\left({r\over R}\right)^3\right\}
\ee
where $B_{0,12}$ is the surface magnetic field in units of $10^{12}$, $r$ is the emission radius and $R$ is
the neutron star radius.  Such cutoffs could be identified by their super-exponential shape, and 
distinguishable from the simple exponential cutoffs due to a maximum in the radiating particle spectrum.  
The detection of a pair attenuation cutoff would require the emission region to be located in the
strong magnetic field within
several stellar radii of the neutron star surface.
Dyks \& Rudak \cite{DR02} have shown that even at relatively low altitudes, aberration and field-line
slippage cause by neutron star rotation can produce asymmetries in pair production cutoffs across the 
polar cap, such that spectra radiated on leading-edge field lines are cut off at lower energies than are
spectra radiated on trailing-edge field lines.  This produces an asymmetry in the pulse profile, as the
trailing peak will dominate at energies approaching the cutoff and the first peak will disappear.
In fact, in the double-peaked profiles of three bright $\gamma$-ray pulsars observed by {\sl EGRET},
the Crab, Vela and Geminga, the trailing peak dominates above 5 GeV \cite{Thomp01}.

It has been suggested \cite{VDJ05} that the high-energy $\gamma$-ray spectra of millisecond pulsars 
may provide constraints on the frame-dragging electric field, discussed in Section \ref{sec:GR}.
Since the electric fields near the surface of millisecond pulsars are expected to be unscreened, due to
insufficient pair production in their very low magnetic fields, the peak of their high-energy 
curvature radiation spectrum, predicted to be at $1 - 50$ GeV, should be a sensitive measure of 
$E_{\parallel}$ \cite{HVM05}.  Non-detection of 
the nearby millisecond pulsar PSR J0437-4715 by {\sl EGRET} already places marginal constraints on the
neutron star equation-of-state dependence in the frame-dragging acceleration model \cite{Harding05}, 
and future observations by air-Cherenkov telescopes will place further constraints.

\section{FUTURE PROSPECTS}

Future measurements by high-energy telescopes that have recently begun operations or are soon to
come will be capable of detecting or significantly constraining most of the signatures that have been 
discussed here.  The {\sl Gamma-Ray Large Area Space Telescope} ({\sl GLAST}) \cite{McE04}, with expected 
launch in 2007, will
have the sensitivity to detect $\gamma$-ray emission between 30 MeV and 300 GeV from possibly several
hundred pulsars.  {\sl GLAST} should easily detect nearby millisecond pulsars like PSR J0437-4715, thereby
putting severe constraints on frame-dragging acceleration models.  Measurement of pulse profiles of 
many pulsars of varying ages and periods, and comparison with their radio profiles, can test the 
predictions of outer gap and slot gap caustic models.  New air-Cherenkov telescopes such 
as {\sl MAGIC} \cite{Lorentz04} and {\sl H.E.S.S.} \cite{Hinton04} 
are expected to achieve sensitivity below 100 GeV and may detect, or at least place further limits on, 
millisecond pulsar spectra.  {\sl INTEGRAL} and {\sl RHESSI} currently have some 
sensitivity to polarized signals at 100-500 keV and $>20$ keV, but not enough for the phase-resolved
polarimetry required to detect signatures of caustic emission in pulsars.
Several proposed X-ray polarimeters are currently under study,
the {\sl Advanced X-Ray Polarimeter} ({\sl AXP}) and the {\sl Polarized Gamma-Ray Observer} ({\sl PoGO}), 
sensitive from 2-10 keV
and 25-200 keV respectively.  Such detectors are expected to have enough sensitivity to distinguish
between the different emission models and to detect caustic signatures in the Crab pulsar.

\newpage
\bigskip 

\end{document}